\documentclass[aps,prl,reprint,floatfix,showpacs]{revtex4-1}
\usepackage{amssymb,amsmath}
\usepackage{graphicx}

\usepackage[scanall]{psfrag}

\usepackage{color}

\newcounter{temp}
\newcommand{\prediction}[1]{\setcounter{temp}{#1}%
								{(\Roman{temp})}%
								}

\graphicspath{{./Fig/fraggedFigs/}}

\begin{document}


\title{Novel self-assembled morphologies from isotropic interactions}%

\author{E. Edlund}%
\author{O. Lindgren}
\author{M. \surname{Nilsson Jacobi}}%
 \email{mjacobi@chalmers.se} 
\affiliation{Complex Systems Group, Department of  Energy and Environment, Chalmers University of Technology, SE-41296 G\"oteborg, Sweden}
\date{\today}%

\begin{abstract} 
We present results from particle simulations with isotropic medium range interactions in two dimensions. At low temperature novel types of aggregated structures appear. We show that these structures can be explained by spontaneous symmetry breaking in analytic solutions to an adaptation of the spherical spin model. We predict the critical particle number where the symmetry breaking occurs and show that the resulting phase diagram agrees well with results from particle simulations.
 \end{abstract}

\pacs{
89.75.Kd, 
75.10.Hk, 
05.65.+b  
}

\maketitle

Understanding the principles behind spontaneous formation of structured morphologies is of interest both as a fundamental scientific question and in engineering applications where the possibility of using self-assembly to produce novel materials provides a compelling complement to traditional blue-printed fabrication~\cite{lehn_toward_2002,zhang_fabrication_2003}. Consequently there is growing interest in exploring interactions that can facilitate self-fabrication of materials with novel properties~\cite{Rechtsman05,chen_directed_2011, osullivan_vernier_2011}. The more general question of what structures are possible to self-assemble from a given class of interactions has however not been addressed, with some notable exceptions e.g.\ simulation based analysis of polyhedral packing~\cite{agarwal_mesophase_2011}. 

To examine the possibilities of self-assembly from a theoretical angle, in this Letter we consider systems with pair-wise isotropic interactions, frequently used as coarse grained models of more complex meso-scale systems such as colloidal systems~\cite{oosawa_surface_1954,*asakura_interaction_1958,*vrij_polymers_1976}, particles in an ambient fluid~\cite{Ercolessi94,Lyubartsev95}, and spin glasses~\cite{klein_statistical_1963}. We show that typical aggregates appearing as low temperature configurations in particle simulations with randomly generated medium range isotropic interactions can be predicted analytically using an adaptation of the spherical spin model~\cite{baxter_exactly_1982}. The morphologies, many of them novel and surprisingly complex, can be systematically classified by their spontaneous breaking of the rotational symmetry.


To formulate a solvable model of a particle system we start by considering a lattice spin system with Hamiltonian on the form 
\begin{equation}
   H = \sum_{ij} U_{ij} s_i s_j,
\label{hamiltonian}
\end{equation}
where $s_i \in \{ 0,1 \}$, and $s_i =1$ represents a particle at lattice site $i$ and $s_i =0$ represents vacuum. The total number of particles is set by the normalization $\sum _i s_i = N$.  The interaction $U_{ij}$ is an effective isotropic potential with a hard core repulsion corresponding to the lattice spacing. In a spin glass metal the interaction is mediated by a polarization of the Fermi sea~\cite{fischer_spin_1993} while in a colloidal system it could involve a surface polymer induced steric hindrance competing with a depletion attraction~\cite{likos_effective_2001}. 

\begin{figure*}[htb]
\includegraphics[width=0.8\textwidth]{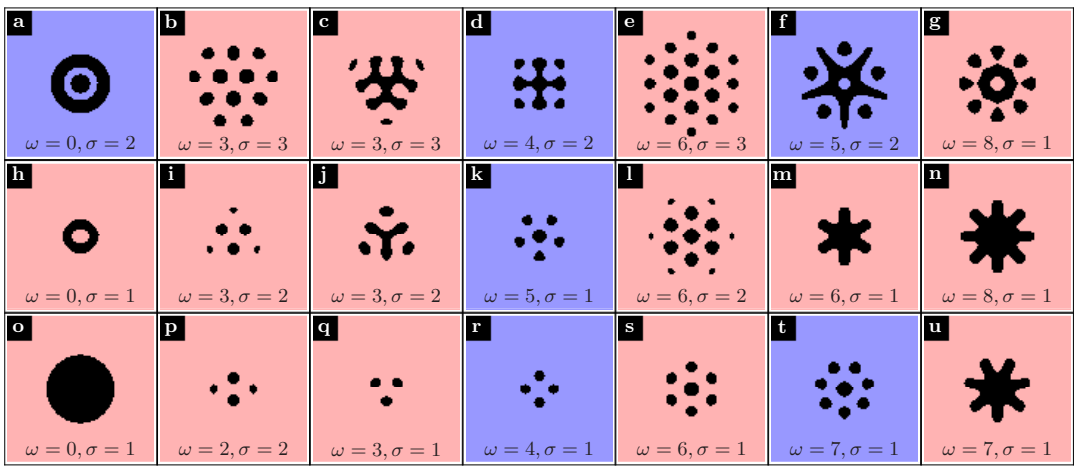}
\caption{ \label{morphology} 
Predicted morphological alphabet for aggregating potentials generated by equation~\eqref{mainResult} (with $\omega \le 8$). The morphologies shown are energetically preferable for a large fraction of random potentials. Red background 
indicates a strong signal while blue background 
signifies weaker signal. $\sigma$ denotes the number of active radial oscillations of the ground frequency after the threshold mapping. 
}
\end{figure*}

The discrete model described by Eq.~\eqref{hamiltonian} can equivalently be formulated as a continuous one ($s_i \in \mathbb{R}$) with auxiliary constraints, $\sum _i s_i ^m = N$ $\forall m$. To make the model analytically tractable we relax the constraints to include only the first two moments. The result is the spherical spin model including an external field, with Hamiltonian $H_s = \sum _{ij} U_{ij} s_i s_j + h \sum _i s_i$.  A necessary condition for an energy minimum in this model is	$\sum _j ( U_{ij} - \lambda \delta _{ij} ) s_j = h/2$, where $\delta $ is the identity matrix, and $\lambda$ comes from the constraint on the second moment. Translational invariance implies that $s_i = c$ is a solution. Non-constant solutions $s_i = v_i + c$ exist if $\lambda$ is an eigenvalue of the matrix $U$ corresponding to the eigenvector $v$. From the constraints it follows that $c = \rho$, where $\rho$ is the density, and $V ^{-1} \sum _i v_i ^2 = (1- \rho ) \rho$, where $V$ is the number of lattice sites or the volume. The ground states are defined by the eigenvector(s) corresponding to the lowest eigenvalue of $U$. 

It can be shown, e.g.\ using translational invariance, that all isotropic interaction matrices have the Fourier normal modes as eigenvectors. This also follows from the observation that any isotropic interaction matrix $U_{ij}$ can be expressed as a linear combination of matrices on the form $\Delta _{+} ^l \Delta _{\times} ^m$, $l,m = 0,1,\dots$, where the two components are the discrete Laplace operator defined as usual $\Delta _+$ or along the diagonals $\Delta _{\times}$ (note that $\Delta _+$ and $\Delta _{\times}$ commute).  The Fourier modes are eigenfunctions of the Laplace operator so the two arguments lead to  the same conclusion. However, the latter argument points to a subtlety: the eigenfunctions of the Laplacian depend on the boundary conditions, where the Fourier harmonics result from periodic boundaries. Requiring that the functions converge to zero at infinity instead results in cylindrical (spherical etc.) harmonic eigenfunctions with angular modulations localized around  a nucleation point. These localized configurations bears similarity to topological defects, such as the Belavin-Polyakov monopole, that are important for the low temperature behavior of e.g. Heisenberg ferromagnets~\cite{Belavin75}.

Our central result is that, when there are too few particles to occupy a translational invariant ground state, the eigenfunctions of Bessel type $J_{\omega}(2\pi k r)\cos{(\omega \theta)}$ determine the low temperature behavior of self-assembling particle systems in two dimensions. 
The eigenvalues, i.e.\ the energies, are given by the radial Fourier transform of the potential as~\cite{edlund_universality_2010}\footnote{The eigenvalues are independent of the eigenbasis and \eqref{eq:spectrum} 
is from~\cite{edlund_universality_2010}, where the Bessel function in the integral has no direct relation to the Bessel eigenfunctions of the Laplacian.}
\begin{equation}
	E ( k ) = 2 \pi \int _0 ^{\infty} r dr \, U(r) J_0 ( 2 \pi k r) ,	
	\label{eq:spectrum}
\end{equation}
so that the wavenumber $\kappa = \operatorname{arg\,min}_{k} E (k)$ defines an eigenfunction with minimal energy. The energy spectrum is degenerate since $E(k)$ is independent of $\omega$, reflecting the rotational invariance around the nucleation point. However, this degeneracy is broken by the non-linearity in the mapping from the spherical model to the discrete lattice system, where a single base frequency $\omega$ together with its overtones (similar to the harmonic overtones of a square wave) are energetically favored. The spontaneous symmetry breaking of the ground state, from $\operatorname{O}(2)$ to one of its isotropy subgroups $D_{\omega}$, is equivalent to the behavior in a bifurcation problem~\cite{Matthews03,Hoyle06}. The ground states of the spherical model of relevance to the self-assembly problem in \eqref{hamiltonian} are therefore on the form
\begin{equation}
\sum_{n=0}^\infty a_n J_{n \omega } ( 2 \pi \kappa r ) \cos (n \omega \theta ) + c J_{0}(2\pi k_m)
\label{mainResult}
\end{equation}
in the limit $k_m\rightarrow 0$, which turns the last term (which we refer to as the mass-builder) into the translational invariant constant $c$ in the minimization of $H_s$. However, the $k_m \rightarrow 0$ limit is only relevant for infinite structures (with non-zero global density). For an aggregated structure with finite mass a small but non-zero $k_m$ is needed to localize the solution (resulting in a non-zero local density but a global density of zero). 

As we will see later, the sum in Eq.~\eqref{mainResult} is dominated by the first few terms and in practice most structures are adequately described by the zeroth and first term. Notable exceptions are finite lattices, see e.g.\ Fig.~\ref{morphology}e, which grow by successively including  higher terms.

\begin{figure*}[htb]
\includegraphics[width=0.95\textwidth]{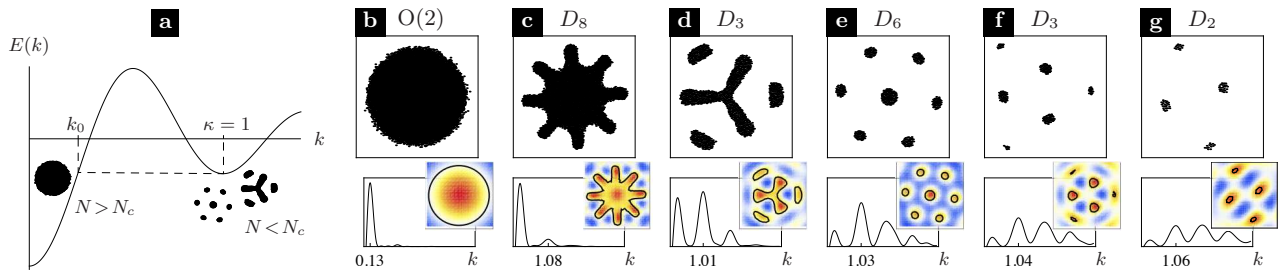}
\caption{ \label{densitySweep} 
Showing self-assembled morphologies for a single potential. (a) A typical energy spectrum of a random potential. The wavelength of the morphologies is determined by the interior minimum $\kappa$ of the spectrum when the number of particles $N$ is less than $N_c$, the number required to build a disk with radius corresponding to $k_0$, see Eq.~\eqref{predictionEq}. (b)-(g) Transitions between different symmetry groups for the potential of (a) as the particle number is decreased. (Top) Configurations from annealed particle simulations, (middle) corresponding states of the spherical model with contours added at a level giving correct number of particles, and (bottom) Bessel spectra (arbitrary units) of the particle configurations expanded on the form of \eqref{mainResult}, summed over $\omega$.  The first maximum corresponds to the mass-builder $k_m$, the second should be compared to $\kappa = 1$ predicted from the spectrum in (a), and the rest are overtones.
}
\end{figure*}



To connect the above results to the particle model \eqref{hamiltonian}, we map the continuous ground states ($s_i \in \mathbb{R}$) to particle configurations ($s_i\in\{0,1\}$) by applying a step function threshold. The energy of the resulting discrete configurations depend on how the mapping distorts the power spectrum relative to the energy spectrum \eqref{eq:spectrum}. While the main weight of the discrete configurations typically remains localized to $\kappa$, contributions from overtones that appears in the discrete system can have a large effect on the energy. These effects are difficult to quantify analytically. Instead we generate possible candidates for ground states using Eq.~\eqref{mainResult}, i.e.\ linear combinations of Bessel functions with angular frequencies $0$, $\omega$,  $2 \omega$ and $3 \omega$ with $\kappa =1$, as well as a mass-building Bessel function with $\omega = 0$ and $k_m$ chosen such that the first zero of the mass-builder is slightly larger than the $\sigma$th zero of $J_\omega$ for $\sigma = 1,2,3$. The configurations were mapped to binary valued configurations on a lattice with a threshold that defines different masses. The energy of the resulting configurations were calculated for $1000$ random first-order spline potentials.  
We used aggregating potentials, i.e.\ potentials with a global minimum in the energy spectrum at $k=0$. In addition their arguments where scaled so that the interior minimum in the energy spectrum resided at $\kappa =1$ as demonstrated in Fig.~\ref{densitySweep}a and were attractive at the lattice distance to ensure correct resolution of the lattice. For each potential and mass we recorded the configuration with the lowest energy, which resulted in a limited number of favored morphologies, shown in Fig.~\ref{morphology}. The method predicts many novel structures (e.g.\ Fig.~\ref{morphology}a,c,d,g,h,m),
as well as simpler ones like disks and localized lattices (e.g.\ Fig.~\ref{morphology}b,e,l,q). The latter are observed on the  atomic scale~\cite{haekkinen_electronic_2003}, while the former morphologies are only expected to appear at higher particle numbers.

To summarize the predictions of the adapted spherical spin model, we expect a particle system with an aggregating potential to show \prediction{1} morphologies from a limited alphabet (Fig.~\ref{morphology}) with \prediction{2} a single dominant wavelength $\kappa$ determined by the minimum of the spectrum and \prediction{3} an $\omega$ degeneracy where a single potential can self-assemble into many different morphologies, depending on external parameters.

\begin{figure}
\includegraphics[width=0.45 \textwidth]{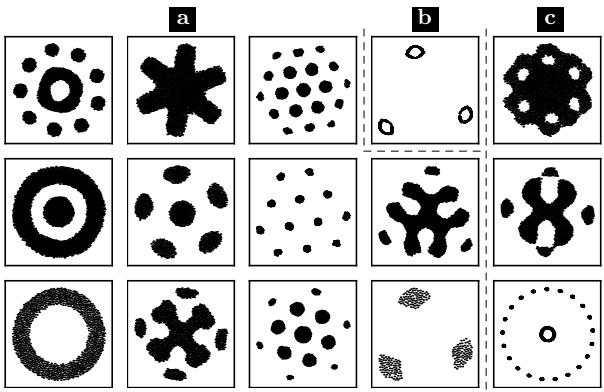}
\caption{ \label{mcConfigs} 
Examples of configurations from particle simulations with randomly generated potentials.
All configurations except (b) are well represented by functions on the form \eqref{mainResult} with the first four terms included, most requiring only one or two. 
(a) Morphologies predicted as common by our method, see Fig.~\ref{morphology}. (b) An example of a hierarchical structure, not directly describable in our theory, where each part of the usual Bessel structure serves as the nucleation point for a separate Bessel structure. 
(c) Morphologies that, while being simply representable by Eq.~\eqref{mainResult}, are not predicted in Fig.~\ref{morphology}. 
}
\end{figure}


To test these predictions we performed off-lattice simulations of particle systems.  We constructed 1200 random interaction potentials (piecewise constant and 3rd order splines with the same restrictions on the spectrum as above) and did Monte Carlo annealing at different particle numbers. Examples of the resulting particle configurations are shown in Fig.~\ref{densitySweep}-\ref{mcConfigs} and are in good agreement with \prediction{1}.
Of the simulated particle systems approximately $86\%$ (Fig.~\ref{mcConfigs}a) annealed to predicted morphologies shown in Fig.~\ref{morphology} and $9\%$ (Fig.\ \ref{mcConfigs}c) annealed to morphologies described by Eq.~\eqref{mainResult} but absent in Fig.~\ref{morphology} due to the limited parameter range considered (e.g.\ $\omega\le 8$). The latter were in general similar to those in Fig.~\ref{morphology}. The frequencies of which predicted configurations where observed was also largely consistent with the strength of the signal of the corresponding morphology in Fig.~\ref{morphology}. 
The remaining $5\%$ of the morphologies exhibited hierarchies of nested Bessel like structures, for an example see Fig.~\ref{mcConfigs}b.

By expressing the observed particle configurations in the modulated Bessel basis we confirm that \prediction{2} the dominant wavenumber is accurately predicted by the minimum in the energy spectrum of the potential. Examples of this are shown in Fig.~\ref{densitySweep}b-g, where we note that while the point symmetry of the preferred configuration for a single potential changes with varying particle number, the dominant (non mass-building) active wavelength is approximately constant and equal to the predicted value $\kappa$. This switching of symmetry group is observed frequently in the simulations and shows that \prediction{3} the $\omega$ degeneracy is not just a mathematical curiosity in our model. For a given potential the particle number sets the morphology, an important fact that could facilitate the use of standard techniques such as density gradient centrifugation~\cite{hinton_density_1978} for differential selection of morphologies~\cite{manoharan_dense_2003}.


\begin{figure}[htb]
\includegraphics[width=0.45 \textwidth]{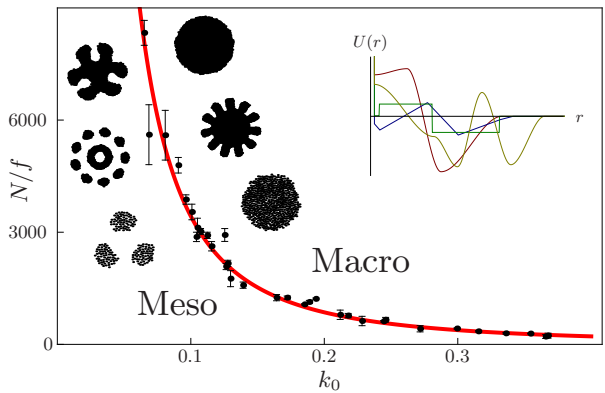}
\caption{ \label{predictionFig} 
Decreasing the number of particles causes a transition from disc-like structures at the macro-scale to more complex morphologies at the meso-scale. The figure shows the phase diagram with the predicted (full line) and measured (markers) critical number of particles $N_c$, adjusted for packing fraction $f$, versus $k_0$ calculated (see Fig.~\ref{densitySweep}a) from respective spectra for 33 random potentials.
(Inset) Four examples of potentials used.
}
\end{figure}

The spectral analysis also allow us to analytically predict a phase diagram for aggregating potentials. The spectrum shown in Fig.~\ref{densitySweep}a is typical for the potentials we consider in that it is oscillatory and has a global minimum at $k=0$. In a spin system a global minimum at $k=0$ implies a ferromagnetic ground state. For a particle system it implies that in the high particle limit we can use large scale interface minimization to argue that the ground state is dominated by the mass-builder, typically a closely packed disc with a surface that is smooth or has wavelike indentations (similar to Fig.~\ref{densitySweep}c). 
For a given number of particles there is a maximal size of such a disc and for low particle numbers this effectively excludes the small $k$ part of the energy spectrum in the minimization that determines $\kappa$. This causes a transition as the particle number is decreased, from a disc to the more complex Bessel based morphologies we observe. This transition happens when the energy of the disc raises above the energy of the lowest interior minimum $E(\kappa)$. We denote this point $k_0$, as illustrated in Fig.~\ref{densitySweep}a. The critical particle number $N_c$ can be predicted through
\begin{equation}
\label{predictionEq}
N_c =  \frac{f}{a^2} \cdot \frac{b^2}{k_0^2}
\end{equation}
where $f$ is the packing fraction, $a$ is the radius of the particles, and $b\approx 0.293$ is a numerical constant corresponding to the optimal $k$ to describe a disc of radius 1, found through $\operatorname*{arg\,max}_k \int_0^1 \!  J_0(2 \pi k r) \, r \mathrm{d}r/||J_0(k)||_2^2$. The phase diagram predicted by Eq.~\eqref{predictionEq} is shown in Fig.~\ref{predictionFig} together with simulation results.


We conclude that isotropic pairwise additive potentials can give rise to complex morphologies appearing between the atomic and macroscopic scale and that the frequently occurring structures forms a limited morphological alphabet. The patterns are generically constructed as discretized linear combinations of a few Bessel functions, which can be understood and analytically predicted from an adaptation of the spherical spin model. Further, we analytically calculate the phase diagram showing where these patterns occur for different potentials. The accuracy of the predictions is surprising considering the complexity involved in determining the ground states in a many particle system. The methodology we present applies to the entire class of isotropic interaction potentials and provides new theoretical understanding of self-assembly processes.

\begin{acknowledgments}
OL and MNJ acknowledge support from the SUMO Biomaterials center. We thank Kolbj\o rn Tunstr\o m for valuable comments and discussions. 
\end{acknowledgments}


%

\end{document}